\documentclass[12pt]{article}
\usepackage{graphicx, url}
\usepackage{palatino}
\usepackage{color}
\usepackage{amsmath, amssymb,bbm}
\usepackage{epsfig}
\usepackage{rotating}
\usepackage{dcolumn}
\usepackage{bm}
\usepackage{caption}
\usepackage{subcaption}
\newcommand{\comment}[1]{}
\def \non{\nonumber}
\def \ra{\rightarrow}
\def \bea{\begin{eqnarray}}
\def \eea{\end{eqnarray}}
\def \Qbar{\overline Q}
\def \qqbar{Q\Qbar}

\def \ccbar{c\overline{c}}

\def \sg{\sigma}

\begin{document}   
\baselineskip 18pt
\title{Resolution of the LHCb {\boldmath $\eta_c$} anomaly}
\author{
   Sudhansu~S.~Biswal$^1$\footnote{E-mail: sudhansu.biswal@gmail.com}, 
   ~Sushree~S.~Mishra$^1$\footnote{Email: sushreesimran.mishra97@gmail.com}
     ~and  K.~Sridhar$^2$\footnote{E-mail: sridhar.k@apu.edu.in} \\ [0.2cm]
    {\it \small 1. Department of Physics, Ravenshaw University,} \\ [-0.2cm]
    {\it \small Cuttack, 753003, India.}\\ [-0.2cm]
    {\it \small 2. School of Arts and Sciences, Azim Premji University,} \\ [-0.2cm]
    {\it \small Sarjapura, Bangalore, 562125, India.}\\
}
\date{}
\maketitle
\begin{abstract}
\noindent Due to the heavy-quark symmetry of Non-Relativistic Quantum Chromodynamics (NRQCD), the
cross-section for the production of $\eta_c$ can be {\it predicted}. This NRQCD prediction when confronted
with data from the LHCb is seen to fail miserably. We address this LHCb $\eta_c$ anomaly in this paper using
a new approach called modified NRQCD, an approach that has been shown to work extremely well for studying
$J/\psi$, $\psi^{\prime}$ and $\chi_c$ production at the LHC. We show, in the present paper, that the
predictions for $\eta_c$ production agrees very well with LHCb measurements at the three different values
of energy that the experiment has presented data for. Modified NRQCD also explains the intriguing agreement
of the LHCb $\eta_c$ data with the colour-singlet prediction. The remarkable agreement of the theoretical
predictions with the LHCb data suggests that modified NRQCD is closer to apprehending the true dynamics of
quarkonium production.
\end{abstract}
\maketitle

\noindent One of the most important and daunting problems in Quantum
Chromodynamics (QCD) 
is the understanding of how quarks form physical bound states 
-- the hadrons. One small corner of this unnavigated terrain where one can make
some headway is in the study of quarkonia -- when a heavy quark and anti-quark
come together to form a neutral meson. A very heavy quark like the top decays
before it can form a bound-state so when we study quarkonia we are interested
in charmonium and bottomonium systems. In these systems, the relative velocity,
$v$, of the $Q \bar Q$ pair is small and, therefore, the bound state can be studied
in a non-relativistic approximation. The effective field theory that has been 
formulated to study such systems is called Non-Relativistic Quantum Chromodynamics
(NRQCD) \cite{bbl} which is derived from the QCD Lagrangian by neglecting all states of momenta much larger 
than the heavy quarkonium mass, $M$ and accounting for this exclusion by adding new interaction 
terms yielding the effective Lagrangian. 

The quarkonium state admits of a Fock-state
expansion in orders of $v$. At leading order, the $\qqbar$ state is in a colour-singlet
state but at {\cal O}$(v)$, it can be in a colour-octet state which is connected to the
physical $J/\psi$ state through a non-perturbative gluon emission.  

The cross section for production of a quarkonium state $H$ of mass $M$ in NRQCD can be 
factorised as:
\bea
  \sigma(H)\;=\;\sum_{n=\{\alpha,S,L,J\}} {F_n\over {M}^{d_n-4}}
       \langle{\cal O}^H_n({}^{2S+1}L_J)\rangle, 
\label{factorizn}
\eea
where $F_n$'s are the short-distance coefficients and ${\cal O}_n$ are 
operators of naive dimension $d_n$, describing the long-distance effects.  
These non-perturbative matrix elements are guaranteed to be
energy-independent due to the NRQCD factorization formula, so that they
may be extracted at a given energy and used to predict quarkonium cross-sections
at other energies.

NRQCD found much success in explaining the systematics of charmonium production at the
Fermilab Tevatron \cite{cdf} in contrast to the then existing model of quarkonium
production -- the colour-singlet model \cite{br}. But NRQCD does not predict the
normalisation of the $p_T$ distributions because of the unknown non-perturbative
parameters so other tests of NRQCD were needed to validate it \cite{tests}. Of these, the polarisation of the
$J/\psi$ provides an important test: NRQCD predicts 
\cite{polar1,polar2} a fully transversely polarised $J/\psi$ at large $p_T$. 
 The Tevatron experiments found no evidence for this \cite{polar3}. 

Another important test of NRQCD comes from the study of $\eta_c$ production.
The heavy-quark
symmetry of NRQCD provides a set of relations which connect non-perturbative
parameters of different resonances so a measurement of a given
state yields information on the non-perturbative parameters of
another state related to the former by heavy-quark symmetry. In particular,
the non-perturbative parameters required for $\eta_c$ production can
be obtained, using heavy quark symmetry, from the parameters of $J/\psi$
production. This approach has been used to predict the $\eta_c$ production cross-section at the Tevatron \cite{Mathews}
and at the LHC \cite{Sridhar:2008sc}. \footnote{$h_c$ production at the Tevatron
\cite{Sridhar:1996vd} and at the LHC \cite{Sridhar:2008sc} has also been studied using this approach}. 

Like the prediction of polarisation, the prediction of the $\eta_c$ cross-section is a definitive test
of NRQCD. Just as NRQCD fails miserably in predicting the $J/\psi$ polarisation, it also gets the
$\eta_c$ cross-section completely wrong \cite{lhcb}: the NRQCD prediction is completely at 
variance with the cross-section measured by the LHCb experiment at three different values of centre-
of-mass energy. It is this LHCb $\eta_c$ anomaly that we address in this paper.

The Fock space expansion of the physical $\eta_c$, which is a $^1S_0$
($J^{PC}=0^{-+}$) state, is: 
\bea
\left|\eta_c\right>={\cal O}(1)
	\,\left|\qqbar[^1S_0^{[1]}] \right>+
	 {\cal O}(v^2)\,\left|\qqbar[^1P_1^{[8]}]\,g \right>+
	{\cal O}(v^4)\,\left|\qqbar[^3S_1^{[8]}]\,g \right>+\cdots ~.
\label{fockexpn}
\eea
In the above expansion the colour-singlet $^1S_0$ state contributes 
at ${\cal O}(1)$. As the $P$-state production is itself down 
by factor of ${\cal O}(v^2)$ both the colour-octet $^1P_1$ and $^3S_1$ 
channels effectively contribute at the same order. The  
colour-octet state $^1P_1^{[8]}$ ($^3S_1^{[8]}$) becomes a 
physical $\eta_c$ by  emitting a gluon in an E1 (M1) transition.  
Keeping terms up-to ${\cal O} (\alpha_s^3 v^7)$ the $\eta_c$ production 
cross section can be written as:
\bea
\sg(\eta_c)&=&
	F_1[^1S_0]\, \left< 0 \right| {\cal O}_1^{\eta_c}
		[^1S_0]\left| 0 \right> \nonumber\\
     && + \frac{F_8[^1P_1]}{M^2}\, \left< 0 \right| {\cal O}_8^{\eta_c}
		[^1P_1]\left| 0 \right> 
        + F_8[^3S_1]\, \left< 0 \right| {\cal O}_8^{\eta_c}
		[^3S_1]\left| 0 \right>, 
\label{etacnrqcd}
\eea
where the coefficients, $F_n$'s, are the cross sections for the production of
$\ccbar$ pair in the respective angular momentum and colour states.

To predict the $\eta_c$ cross-section, the values of  
$\left< {\cal O}_n^{\eta_c}\right>$'s obtained from the experimentally 
predicted values $\langle {\cal O}_n^{J/\psi}\rangle$'s using relations
given by heavy-quark symmetry:
\bea
\left< 0 \right|{\cal O}_1^{\eta_c}[^1S_0]\left| 0\right>&=&
{1 \over 3} \left< 0 \right|{\cal O}_1^{J/\psi}[^3S_1]\left| 0 \right>
\;(1+ O(v^2)), \non\\
\left< 0 \right|{\cal O}_8^{\eta_c}[^1P_1]\left| 0\right>&=&
\left< 0 \right|{\cal O}_8^{J/\psi}[^3P_0]\left| 0 \right>
\;(1+ O(v^2)), \non\\
\left< 0 \right|{\cal O}_8^{\eta_c}[^3S_1]\left| 0\right>&=&
\left< 0 \right|{\cal O}_8^{J/\psi}[^1S_0]\left| 0 \right>
\;(1+ O(v^2)).
\label{Ovalues}
\eea

With the values of the non-perturbative parameters determined as above, the
$\eta_c$ cross-section is a prediction of NRQCD that can be directly tested
in experiments. The $p_T$-dependence of the $\eta_c$ cross-section has been 
measured at the LHCb at three different energies. In Fig. 1, we compare the predictions
of NRQCD obtained, using the heavy-quark symmetry relations, with the $\eta_c$
$p_T$ distribution at $\sqrt{s}=13$ TeV. The disagreement between NRQCD predictions
and the experimental values of the cross-section is huge. The other conundrum is
that the colour-singlet prediction alone seems to be well in agreement with the
data in total contrast to the situation with $J/\psi$ production. The results
presented in Fig. 1 are not new: this anomaly, as we mentioned earlier, has been known and noted in the
literature for several years now \cite{lhcb}; we have presented the results in
Fig. 1 to draw attention to the magnitude of the discrepancy and to also
bring to the fore the mysterious agreement of the colour-singlet prediction
with the data.

\begin{figure}[h!]
\begin{center}
\includegraphics[width=11cm]{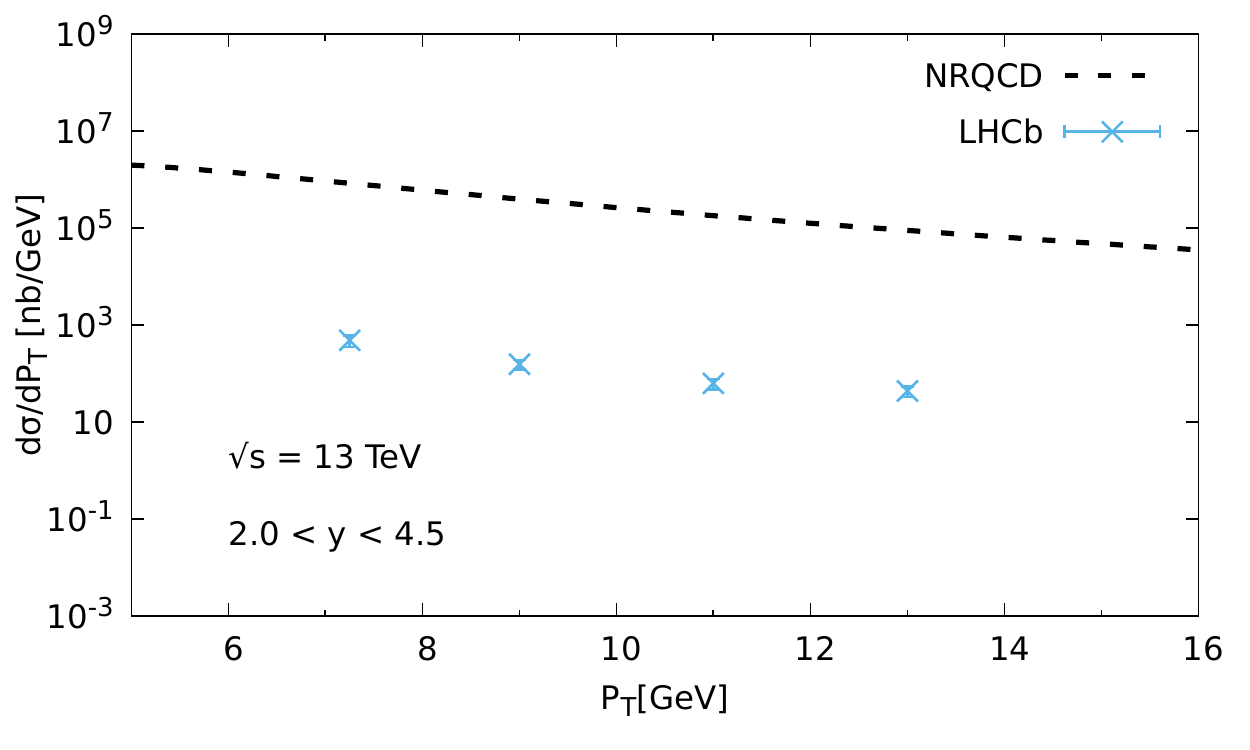}
\caption{ NRQCD predicted differential cross section 
compared with the data on $\eta_c$ production from the LHCb experiment
 at $\sqrt{s} = 13$ TeV. }
\label{fig:1}
\end{center}
\end{figure}

In a recently proposed modification of NRQCD \cite{bms}, which we have named modified NRQCD, we had suggested that
the colour-octet $c \bar c$ state can radiate several 
soft {\it perturbative} gluons -- each emission taking away little energy but carrying 
away units of angular momentum. In the multiple emissions that the colour-octet state
can make before it makes the final NRQCD transition to a quarkonium state, the
angular momentum and spin assignments of the $c \bar c$ state changes constantly. 
Consequently, the Fock expansion for $J/\psi$ (analogous to the one given in Eq.~\ref{fockexpn} for $\eta_c$) 
is no longer valid in modified NRQCD and the equation for $J/\psi$ corresponding to Eq.~\ref{etacnrqcd}
also gets changed.

In Refs.~\cite{bms} and \cite{bms2}, we have studied $J/\psi$ and $\chi_c$ production, respectively,
in modified NRQCD. Fitting the non-perturbative parameters from the Tevatron $J/\psi$ and $\chi_c$ data,
we have made predictions for the cross-sections of these charmonium states at the LHC and find excellent
agreement with the LHC data.

In this paper, we confront the LHCb $\eta_c$ cross-section using modified NRQCD and we present
the details in the following.

For $\eta_c$, for example, the NRQCD cross-section formula which was given as follows when
written down explicitly in terms of the octet and singlet states
\begin{eqnarray}
\sigma_{\eta_c}  = \hat F_{{}^{1}S_0^{[1]}} \times \langle {\cal O} ({}^{1}S_0)^{[1]}) \rangle  + {1 \over M^2} \biggl\lbrack\hat F_{{}^{1}P_1^{[8]}} \times \langle {\cal O} 
                     ({}^{1}P_1)^{[8]} \rangle \biggr\rbrack+
                \hat F_{{}^{3}S_1^{[8]}} \times \langle {\cal O} ({}^{3}S_1)^{[8]}) \rangle  
\label{Fock}
\end{eqnarray}
gets modified to the following in the modified NRQCD with perturbative soft gluon emission:
\begin{eqnarray}
\sigma_{\eta_{c}} &=& \biggl\lbrack \hat F_{{}^{1}S_0^{[1]}} 
		\times ({{\langle {\cal O}^{J/\psi} ({}^{3}S_1^{[1]}) } \rangle \over 3} )\biggr\rbrack \cr 
                &+& \biggl\lbrack  
                  \hat F_{{}^{3}S_1^{[8]}} 
                 + \hat F_{{}^{1}P_1^{[8]}} 
                + \hat F_{{}^{1}S_0^{[8]}} + \hat F_{{}^{3}P_J^{[8]}}  \biggr\rbrack 
		\times ({{\langle {\cal O}^{J/\psi} ({}^{3}S_1^{[1]}) } \rangle \over 8} ) \cr
                &+& \biggl\lbrack  
                  \hat F_{{}^{3}S_1^{[8]}} 
                 + \hat F_{{}^{1}P_1^{[8]}} 
                + \hat F_{{}^{1}S_0^{[8]}} + \hat F_{{}^{3}P_J^{[8]}}  \biggr\rbrack 
		\times \langle {\cal O}^{\eta_c}  \rangle ,
\label{modified1}
\end{eqnarray}
where
\begin{equation}
	\langle {\cal O}^{\eta_c}  \rangle =
                \times \biggl\lbrack 
                 \langle {\cal O} ({}^{3}S_1^{[8]}) \rangle 
                + \langle {\cal O} ({}^{1}S_0^{[8]}) \rangle 
                + {\langle {\cal O} ({}^{3}P_J^{[8]}) \rangle \over M^2}
                    \biggr\rbrack.  
\end{equation}

It is important to pay attention to the second line in Eq.~\ref{modified1}. This term
arises because of the following physical situation in modified NRQCD: in emitting soft gluons, the
colour-octet state can every once in a while make a transition to a colour-singlet state and this
colour-singlet-state can no longer emit any soft-gluons but will eventually make a non-perturbative
transition to a physical charmonium state. This is a mixed octet-singlet contribution and there is nothing 
like this in NRQCD: it is a novel feature of modified NRQCD.

The non-perturbative parameters for $\eta_c$ are not obtained from any independent fits but from the
$J/\psi$ case, using the heavy-quark symmetry relations alluded to below. These are:

\bea
\left< {\cal O}_1^{\eta_c}[^1S_0] \right>&=&
{1 \over 3} \left< {\cal O}_1^{J/\psi}[^3S_1] \right>, 
\\
\left< {\cal O}^{\eta_c} \right>&=& 
\left< {\cal O}^{J/\psi} \right>, 
\label{Ovalues2}
\eea
where $\left< {\cal O}^{J/\psi} \right>$ 
is the fitted parameter for $J/\psi$. We have taken 
$\left< {\cal O}^{J/\psi} \right> = - 0.161~ \rm{GeV}^3$,  
which was obtained earlier using modified NRQCD~\cite{bms}.

The cross-section kinematics are the usual and, even at the risk of pedantry, we write the expression
for the cross-section explicitly:
\bea
&&\frac{d\sg}{dp_{_T}} \;(p \bar p \ra \ccbar\; [^{2S+1}L^{[1,8]}_J]\, X)= \non \\
&&\sum \int \!dy \int \! dx_1 ~x_1\:G_{a/p} (x_1)~x_2\:G_{b/p}(x_2) 
\:\frac{4p_{_T}}{2x_1-\overline{x}_T\:e^y}\non\\
&&\frac{d\hat{\sg}}{d\hat{t}}
(ab\ra \ccbar[^{2S+1}L_J^{[1,8]}]\;d),
\label{eq:diff}
\eea
where the summation is over the partons ($a$ and $b$),    
$G_{a/p}$,   
$G_{b/p}$ are the distributions of partons $a$ and $b$ in the
protons and $x_1$, $x_2$ are the respective  momentum they carry.
In the above formula, $\overline{x}_T=\sqrt{x_T^2+4\tau} \equiv 2 M_T/\sqrt{s}$ \ with \  
$x_T=2p_{_T}/\sqrt{s}$ and  \(\tau=M^2/s\).
$\sqrt{s}$ is the center-of-mass energy, 
$M$ is the mass of the resonance and $y$ is the rapidity at which the 
resonance is produced. 
The matrix  elements for the subprocesses can be found in Refs.~\cite{cho2,gtw, Mathews}. 

\begin{figure}[h!]
\begin{center}
\includegraphics[width=15cm]{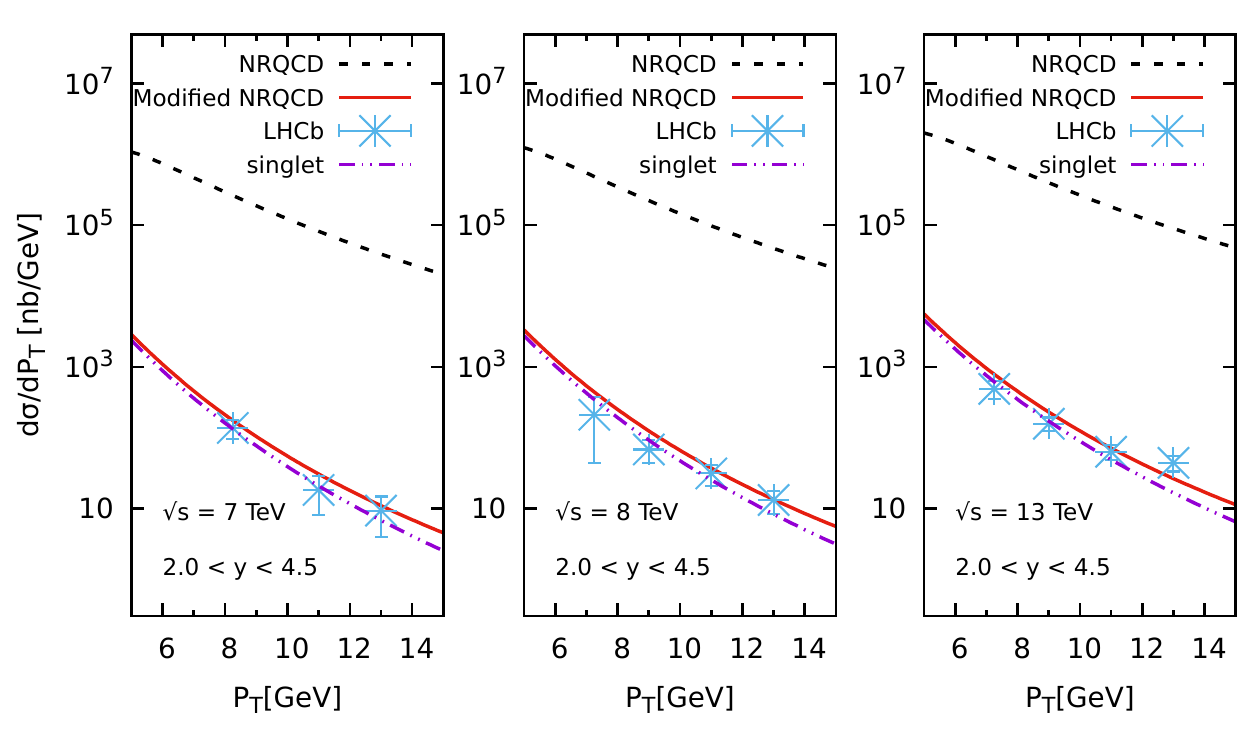}
        \caption{Predicted differential cross sections for $\eta_c$ production at the LHC 
 running at different center-of-mass energies 
	compared with the data from the LHCb experiment.}
        \label{fig:fig2}
\end{center}
\end{figure}

In Fig.~2 we compare the predictions for the $\eta_c$ cross-section in modified NRQCD with the 
experimental measurements at different energies from the LHC experiment. The remarkable agreement
of the predictions with LHCb cross-sections are obvious and the agreement is seen at all the energies.
%
%
%
We have also shown the colour-singlet prediction separately in Fig.~2. As can be seen,
the modified NRQCD prediction and the colour-singlet prediction are very similar. In other words, in
modified NRQCD, there is a delicate contribution between the octet contribution and the mixed octet-singlet
contribution and that is why the total contribution is essentially given by the colour-singlet contribution.
we should add that this cancellation materialises only in the forward kinematic region (where LHCb
measures the $\eta_c$ cross-section) and not in the central region where the LHC experiments or even
the Tevatron experiments measured their $J/\psi$ cross-section. 

\begin{figure}[h!]
\centering
\includegraphics[width=11cm]{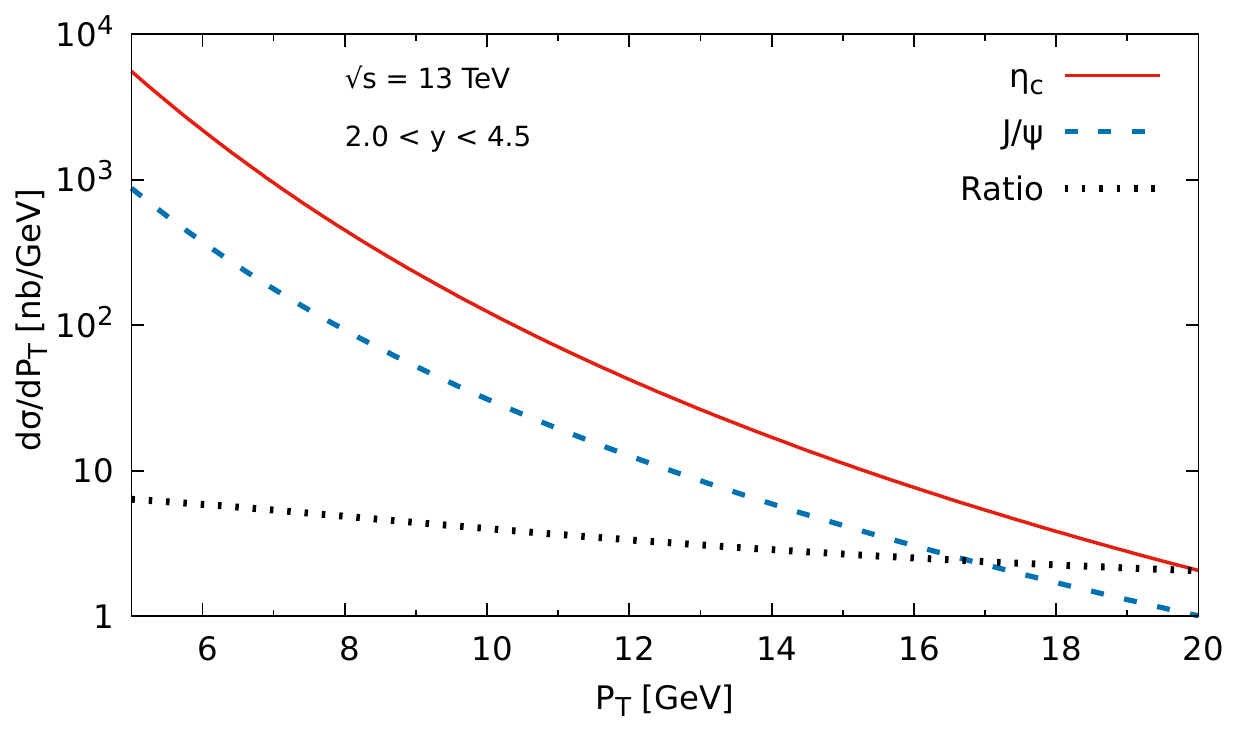}
\caption{Ratio of $\eta_c$ to $J/\psi$ differential production cross-sections at the LHC for $\sqrt{s} = 13$ TeV.}
\label{fig:fig3}
\end{figure}

Finally, we present in Fig.~3 the ratio of the $\eta_c$ to $J/\psi$ cross-section at 
$\sqrt{s}=13$ TeV and in the same forward region of kinematics, a measurement that may be undertaken in
the LHCb experiment in the near future.

In summary, modified NRQCD provides a neat solution to the LHCb $\eta_c$ anomaly and provides an understanding
of all the features of the $\eta_c$ data. It is important to reiterate that the $\eta_c$ cross-section in modified
NRQCD is a prediction and not a fit and the remarkable agreement with the LHCb data suggests that modified NRQCD
is closer to apprehending the true dynamics of quarkonium production.
%

\section*{Acknowledgments} One of us (K.S.) is grateful to members of the LHCb collaboration -- Monica Pepe-Altarelli,
Sergey Barsuk, Valeriia Zhovkovska and Andrii Usachov for valuable discussions. Discussions with Vaia Papadimitriou
are also gratefully acknowledged.


\end{document}